\documentclass[twocolumn,aps,showpacs,preprintnumbers,amsmath,amssymb]{revtex4}
\usepackage{epsfig}
\usepackage{amssymb}

\begin{document}

\title{A Quantum Optical Spring}

\author{ Amit Rai and G. S. Agarwal} \affiliation{Department of Physics,
Oklahoma State University, Stillwater, Oklahoma 74078, USA}

\date{\today}

\begin{abstract}
We study the dynamics of the quantum optical spring, i.e., a
spring whose spring constant undergoes discreet jumps depending on
the quantum state of another system. We show the existence of
revivals and fractional revivals in the quantum dynamics
reminiscent of similar dynamical features in cavity QED. We
recover in the semi classical limit the results for an oscillator
whose frequency undergoes a sudden change. The quantum optical
spring is conceivable for example by a micromirror under the
influence of radiation pressure by a field which is strictly
quantum. Our work suggests that driven systems would in general
exhibit a very different dynamics if the drive is replaced by a
quantum source.
\end{abstract}


\maketitle\section{Introduction}The quantum dynamics of an
oscillator with time dependent modulation of its frequency has
been investigated very extensively \cite{janz,gsa,Graham,Glauber}.
This is because of its many applications in different contexts.
One very interesting application was in the context of center of
mass motion of an ion in Paul trap. More recently the study of
modulated oscillator acquired new importance in connection with
BEC's where trap potential can easily be manipulated
\cite{summy,biegelow}. Further nanomechanical oscillators provided
new impetus for the study of quantum dynamics of frequency
modulated system \cite{blencowe}. One of the important outcome of
such studies was that mechanical and vibrational degrees of
freedom can be prepared in nonclassical states \cite{gsa,garrett}.
In particular, one found that sudden changes in the frequency of
the oscillator yield significant amount of squeezing.

\noindent In all the above studies the oscillator was treated as a
quantum system whereas the source of modulation was considered as
something prescribed. Clearly a complete quantum dynamical theory
should consider the source also to be quantized system. This is
important as the mechanical oscillator can have significant back
action on the source of modulation. The system is then a quantum
optical spring , i.e., a spring whose spring constant depends on
the quantum state of another system. The modulation by a quantized
source produces new features in the quantum dynamics as the
oscillator samples the discrete structure of the quantized source
of modulation. In particular we show the appearance of
characteristic quantum revivals and fractional revivals
\cite{Eberly,Schleich,Leichtle} . The dynamics we present should
be realized by the effects of radiation pressure on micromirror
\cite{arcizet,footnote1}.

The organization of the paper is as follows : In section II, we
describe the model and derive analytical result for the reduced
density matrix of the oscillator and the source of modulation. The
quantum dynamics of the oscillator is discussed in section III. In
section IV, we analyze the squeezing properties. The back action
effect is described in section V.

\section{MODEL FOR THE QUANTUM OPTICAL SPRING}

Consider a harmonic oscillator with mass $m$ and frequency
$\omega$. Its Hamiltonian is given by

\begin{eqnarray}
 H_0&=& \frac{p^2}{2m}+\frac{m \omega^{2} x^2}{2}~,
 \label{eq1}
\end{eqnarray}

\noindent where $x$ and $p$ are the position and momentum
coordinates obeying the canonical commutation relation $[x,p]=i
\hbar$ . Let us assume that the frequency of the oscillator is
modulated by a quantized source such that the Hamiltonian in Eq.
(\ref{eq1}) goes to

\begin{eqnarray}
 H &=&\frac{p^2}{2m}+\frac{m \omega^{2}(1+\mu a^\dagger a)
 x^2}{2}~.
 \label{eq2}
\end{eqnarray}

\noindent Since the interaction part commutes with the unperturbed
Hamiltonian of the quantized source we do not write this part of
the energy. Further for brevity we would denote by QSM as the
quantized source of modulation. The operators $a$ and $a^\dagger$
for the QSM obey the Boson commutation relations
$[a,a^\dagger]=1$. The Hamiltonian (\ref{eq2}) describes the
quantum optical spring. Let the eigenstates and eigenvalues
$E_{n}^{(0)}$ of $H_{0}$ be denoted by

\begin{eqnarray}
&& \phi_{n}(x) = N_n H_n(\alpha
x)\exp{(-\frac{1}{2}\alpha^2 x^2)},\nonumber\\
&&
E_{n}^{(0)}=\hbar \omega\big(n+\frac{1}{2}\big),\nonumber\\
&&N_{n} = {\big(\frac{\alpha}{\sqrt{\pi} 2^{n}
{n}!}\big)}^\frac{1}{2},\nonumber\\
&&\alpha \equiv \big(\frac{m
{\omega}}{\hbar}\big)^\frac{1}{2}~.\label{eq3}
\end{eqnarray}

\noindent where $H_n$ is the $n$th Hermite polynomial.

\noindent The eigenstates of $H$ are given by

\begin{eqnarray}
&& a^\dagger a |p\rangle = p |p\rangle,\nonumber\\
&&
H \psi_n^{p}|p\rangle = E_{n}^{(p)} \psi_n^{p}|p\rangle,\nonumber\\
&& E_{n}^{(p)} = \hbar\omega_p(n+\frac{1}{2}),\nonumber\\
&&\omega_{p} \equiv \omega \sqrt{(1+\mu p)}~.\label{eq4}
\end{eqnarray}

and where the wave function $\psi_n^{p}$ is given by \noindent

\begin{eqnarray}
&& \psi_{n}^{p} = N_{n} H_n(\alpha_p
x)(\sqrt{\frac{\alpha_p}{\alpha}})\exp{(-\frac{1}{2}\alpha_p^2 x^2)},\nonumber\\
&&\alpha_p \equiv \big(\frac{m
{\omega_p}}{\hbar}\big)^\frac{1}{2}~.\label{eq3}
\end{eqnarray}

\noindent Note that $\psi_n^{p}$ is the eigenstate of harmonic
oscillator with frequency $\omega$ replaced by $\omega_p$. For a
fix $p$, these form a complete set.

\noindent The time evolution of the initial state can be obtained
from the knowledge of the states given by Eq. (\ref{eq4}). Let us
consider the initial state of the coupled system given by

\begin{eqnarray}
\psi(t=0) &=& \sum_{p,n}
C_{p\hspace{0.01cm}n}\phi_{n}(x)|p\rangle~,
 \label{eq6}
\end{eqnarray}

\noindent where $C_{p\hspace{0.01cm}n}$ are the expansion
coefficients. The state at time $t$ would then be

\begin{eqnarray}
|\psi(t)\rangle
 & = &
 \exp({\frac{-i H t}{\hbar}})\psi(t=0)\nonumber\\
& = & \sum C_{p\hspace{0.01cm}n} |p\rangle
\exp[{\frac{-i t}{\hbar}}(\frac{p^2}{2m}+\frac{m \omega_{p}^{2} x^2}{2})]{\phi_{n}(x)}\nonumber\\
& &\nonumber\\
& = &
 \sum C_{p\hspace{0.01cm}n} |p\rangle
\exp[{\frac{-i t}{\hbar}}(\frac{p^2}{2m}+\frac{m \omega_{p}^{2}
x^2}{2})]\nonumber\\
& & \times  \sum_{l}
|\psi_l^{p}\rangle{\langle\psi_l^{p}|\phi_{n}\rangle} \nonumber\\
& = &
 \sum_{p, n,l}C_{p\hspace{0.01cm}n}\exp[{\frac{-i
t E_{l}^{p}}{\hbar}}]{\langle\psi_l^{p}|\phi_{n}\rangle} |p\rangle
|\psi_l^{p}\rangle ~. \label{eq7}
\end{eqnarray}

\noindent This is the state of the combined system of oscillator
and the QSM. It is a nonfactorized state and thus the spring gets
entangled to the quantum system controlling the spring constant.
The reduced state of the oscillator and QSM can be obtained by
projecting out the degrees of freedom of the other system. It is
seen that the density matrix for the oscillator is

\begin{eqnarray}
 \rho_{o} &=& \sum_{n, l, m, j, p} |\psi_l^{p}\rangle \langle\psi_j^{p}|  C_{p\hspace{0.01cm}n}C^{*}_{p\hspace{0.01cm}m}  \nonumber\\
& & \times \exp[\frac{-i t (E_{l}^{p}-E_j^{p})}{\hbar}
]{\langle\psi_l^{p}|\phi_{n}\rangle}{\langle\phi_m|\psi_j^{p}\rangle}~,\label{eq8}
\end{eqnarray}

and that for QSM is

\begin{eqnarray}
 \rho_{QSM} &=& \sum_{p_{1}, p_{2}}d_{p_1 p_2}|p_{1}\rangle \langle
 p_{2}|~,
 \label{eq9}
\end{eqnarray}

\begin{eqnarray}
 d_{p_1 p_2} &=& \sum_{n, l, m, j}C_{p_{1}\hspace{0.01cm}n}C^{*}_{p_{2}\hspace{0.01cm}m}{\langle\psi_{j}^{p_{2}}|\psi_l^{p_{1}}\rangle} \nonumber\\
& & \times {\langle \psi_l^{p_{1}}|\phi_n\rangle} {\langle\phi_{m}
|\psi_{j}^{p_{2}}\rangle} \exp[\frac{-i t
 (E_{l}^{p_{1}}-E_j^{p_{2}})}{\hbar}]~.\nonumber\\
\label{eq10}
\end{eqnarray}

\noindent Note specially the rather involved form of the reduced
density matrix for the QSM. This is because the scalar product of
the wave functions $\psi_{l}^{p}$ for different $p$ is nonzero-

\begin{eqnarray}
 {\langle \psi_{j}^{p_{2}}|\psi_l^{p_{1}}\rangle} &\neq & 0 \hspace{0.1cm} \text{for $p_{1}\neq p_{2}$}\nonumber\\
&=& \delta_{l j}\hspace{0.1cm} \text{if $p_{1}=p_{2}$}~.
 \label{eq11}
\end{eqnarray}

\noindent Using Eq. (\ref{eq8}) we can study the details of the
quantum dynamics of the oscillator coupled to QSM. We would like
to mention that the quantum characteristics of a nanomechanical
oscillator coupled to a quantized photon source has been studied
by several authors \cite{marshall,jacobs,armour,law} who have
shown how nonclassical states like CAT states of such a system can
be generated. The Hamiltonian used by previous authors is given by
$a^\dagger a \hspace{0.02cm}x$ which is linear in the oscillator
variable and is different from $a^\dagger a \hspace{0.02cm}x^2$
which is produced by modulation. However Arcizet et al.
\cite{arcizet} have used precisely this Hamiltonian as they argue
that the radiation pressure leads to modulation of the spring
constant. Another possibility to realize quantum optical spring
would be the center of mass motion in a detuned cavity.

\section{QUANTUM DYNAMICS OF THE QUANTUM OPTICAL SPRING : REVIVALS AND FRACTIONAL REVIVALS}

Let us consider first a simple case when the QSM is prepared in a
coherent state with mean amplitude $\alpha$ and the oscillator is
prepared in its ground state. Thus the coefficients $C_{pn}$ are
given by

\begin{eqnarray}
C_{p n} & = & \delta_{n 0}\frac {\alpha^p
\exp(\frac{-|\alpha|^2}{2})}{\sqrt{(p)!}}~. \label{eq12}
\end{eqnarray}

\noindent Using Eq. (\ref{eq12}) in Eq. (\ref{eq8}) and on
simplification we get

\begin{eqnarray}
 \rho_{o} &=& \sum  \frac {|\alpha|^{2p}
\exp({-|\alpha|^2})}{{(p)!}}\exp[{-i t
 \omega_{p}(l-j)}] {|\psi_l^{p}\rangle}{\langle
 \psi_{j}^{p}|}\nonumber\\
& &\times
 {\langle\psi_{l}^{p}|\phi_{0}\rangle}{\langle\phi_{0}|\psi_{j}^{p}\rangle}~.\nonumber\\
 \label{eq13}
\end{eqnarray}

\noindent And in particular the probability of finding the
oscillator in the initial state is given by

\begin{eqnarray}
P_{0}(t) &=& \langle \phi_{0}|\rho_{o} |\phi_{0}\rangle\nonumber\\
&=& \sum _{p}{(|\sum _{l}\exp[-i t \omega_{p}\hspace{0.01cm} l]
{|{\langle\psi_{l}^{p}|\phi_{0}\rangle}|}^2|)}^2 \hspace{0.04cm}\nonumber\\
& & \times\left(\frac {|\alpha|^{2p}
\exp(-|\alpha|^2)}{{(p)!}}\right)\nonumber\\
&=& \sum _{p}{|A_{p}|}^2 \frac {|\alpha|^{2p}
\exp(-|\alpha|^2)}{{(p)!}}~,\label{eq14}
\end{eqnarray}

\noindent where

\begin{eqnarray}
A_{p} &=& \sum_{l}\exp[-i t \omega_{p} l]
{|{\langle\psi_{l}^{p}|\phi_{0}\rangle}|}^2~. \label{eq15}
\end{eqnarray}

\noindent This is one of our key results. The quantum state of the
source appears through the weight factor in Eq. (\ref{eq14}) and
through the discrete sum in Eq. (\ref{eq14}). Such sums have been
encountered before in connection with the cavity QED problems
\cite{Eberly,Schleich} and have been experimentally studied
\cite{Haroche,walther}. The sum in Eq. (\ref{eq15}) can be
evaluated in closed form since

\begin{eqnarray}
{{\langle\psi_{l}^{p}|\phi_{0}\rangle}} &=& \int \psi_{l}^{p}(x)
\phi_{0}(x) dx\nonumber\\
& = &  N_{l} N_{0}(\sqrt{\frac{\alpha_p}{\alpha}}) \nonumber\\
& & \times \int H_l(\alpha_p x)
\exp[(-\frac{1}{2}(\alpha_p^2+\alpha^2)
x^2)] dx \nonumber\\
& =
& N_{l} N_{0}({\frac{\beta_{p}}{\sqrt{\alpha_p \alpha}}}) \nonumber\\
& & \times \int H_l(\beta_p y) \exp[-y^2] dy \nonumber\\
& = & N_{2m} N_{0}({\frac{\beta_{p}}{\sqrt{\alpha_p \alpha}}})
[\frac{\sqrt{\pi}({2m})!(\beta_{p}^2-1)^m}{(m)!}]\nonumber\\
& = & \frac{\beta_{p}}{(\eta_p)^\frac{1}{8}}
[\frac{\sqrt{(2m)!}(\beta_{p}^2-1)^m}{2^m (m)!}]~,
 \label{eq16}
\end{eqnarray}

\noindent where
\begin{eqnarray}
\eta_{p} &=&(1+\mu p)~, \nonumber\\{\beta_p}^2 & = &
\frac{2(\eta_p)^\frac{1}{2}}{{1+(\eta_p)^\frac{1}{2}}}~.
\label{eq17}
\end{eqnarray}

\noindent On using Eq. (\ref{eq16}) in Eq. (\ref{eq15}) we find
closed form expression for $A_{p}$ :

\begin{eqnarray}
A_{p} &=&
\frac{|\beta_p|^2}{\eta_p^\frac{1}{4}(1-(\beta_p^2-1)^2\exp(-2
i\omega_pt))^\frac{1}{2}}~, \label{eq18}
\end{eqnarray}

\noindent and $P_{0}(t)$ can be written as

\begin{eqnarray}
P_{0}(t) &=& \langle|A_{p}|^2\rangle~, \label{eq19}
\end{eqnarray}

\noindent and where $\langle \rangle$ is the average over the
distribution of $p$.

\noindent For the classical source of modulation the corresponding
result would be $|A_{p}|^2$ with $p$ replaced by the strength of
modulation $|\alpha|^2$

\begin{eqnarray}
{{P_{c l}(t)}} &=&
\frac{|\beta_{\alpha}|^4}{\sqrt{\eta_{\alpha}(1-2
(\beta_{\alpha}^2-1)^2\cos(2\omega_{\alpha}
t)+(\beta_{\alpha}^2-1)^4)}}~,\nonumber\\
\eta_{\alpha}& = &  (1+\mu |\alpha|^2)~,\nonumber\\ & &
\nonumber\\{\beta_{\alpha}}^2 & = &
\frac{2(\eta_{\alpha})^\frac{1}{2}}{1+(\eta_{\alpha})^\frac{1}{2}}~, \nonumber\\
\omega_{\alpha} & = & \omega \sqrt{(1+\mu |\alpha|^2)}~.
\label{eq20}
\end{eqnarray}

\begin{figure}
 \scalebox{0.75}{\includegraphics{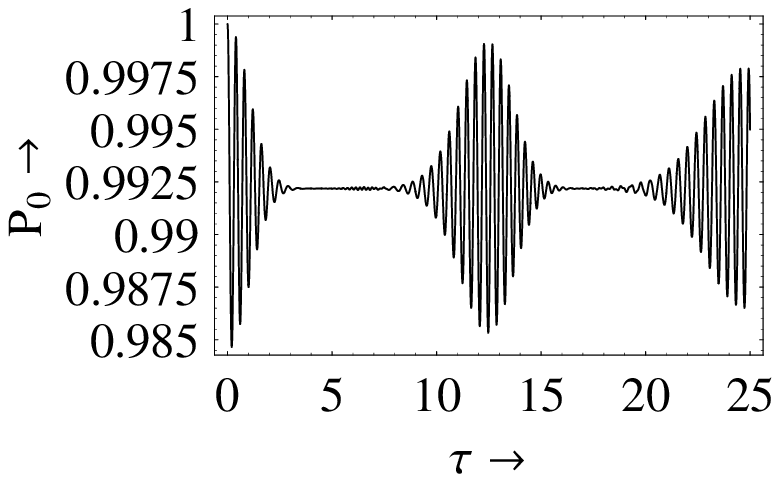}}% Here is how to import EPS art
 \caption{\label{Fig2(a)}Figure shows the variation of $P_{0}$ as a
function of $\tau(\omega t=2 \pi \tau)$ with parameter $\mu=0.1$.
The average number $|\alpha|^2$ of excitation in the QSM is 4.}
 \end{figure}

\begin{figure}
 \scalebox{0.75}{\includegraphics{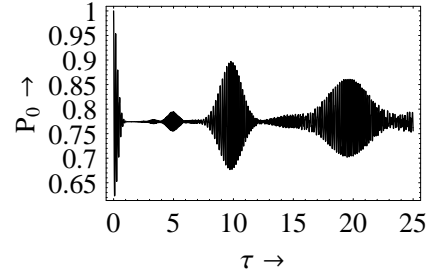}}% Here is how to import EPS art

\caption{\label{Fig2(a)}Figure shows the variation of $P_{0}$ as a
function of $\tau(\omega t=2 \pi \tau)$ with parameter
$\mu=0.3$.The average number $|\alpha|^2$ of excitation in the QSM
is 25.}
 \end{figure}

\noindent Thus $P_{c\hspace{0.05cm}l}$ oscillates at the frequency
$2\hspace{0.08cm}\omega_{\alpha}$. However the probability
$P_{0}(t)$ for the QSM behaves quite differently as shown in the
Fig. 1 and exhibits collapse and revival of the classical periodic
motion. In a different parameter regime the fractional revivals in
$P_{0}(t)$ are seen in Fig. 2. Clearly the dynamics of the quantum
optical spring is very sensitive to the quantized source of
modulation.

\section{SQUEEZING PROPERTIES OF QUANTUM OPTICAL SPRING}

We next investigate if the oscillator system acquires important
nonclassical character as a result of QSM. As mentioned earlier
even a classical change in the frequency can lead to squeezing in
oscillator and therefore an interesting question would be how this
squeezing character is modified by the QSM. In order to calculate
squeezing it is more convenient to work with Heisenberg equations
of motion. For the Hamiltonian in  Eq. (\ref{eq2}) we can prove
that

\begin{eqnarray}
 \frac{d(a^\dagger a)}{dt} &=& 0~,
\end{eqnarray}

\begin{eqnarray}
 \frac{dx}{dt} &=& \frac{p}{m}~,
\end{eqnarray}

\begin{eqnarray}
 \frac{dp}{dt} &=& -m (\omega)^2(1+\mu a^\dagger a )x~,
\end{eqnarray}

\noindent Thus $a^\dagger a$ is a constant of motion and we can
write the solution as

\begin{eqnarray}
x(t) &= & x(0)\cos(\hat{\Omega}t)+\frac{p(0)}{m
 \hat{\Omega}}\sin(\hat{\Omega}t)~,\label{eq22}
\end{eqnarray}

\begin{eqnarray}
p(t) &= & p(0)\cos(\hat{\Omega} t)-m \hat{\Omega}\hspace{0.08cm}
x(0)
 \sin(\hat{\Omega}t)~,\label{eq23}
\end{eqnarray}

\noindent where $\hat{\Omega}$ is given by

\begin{eqnarray*}
\hat{\Omega} &= &\omega \sqrt{1+\mu a^\dagger a}\nonumber\\
\end{eqnarray*}

\noindent and where $a^\dagger a$ is the operator at time $t=0$.

\noindent We define the squeezing variables
\begin{eqnarray}
V_x(t)=\frac{\langle x^2(t)\rangle- {\langle x(t)\rangle}^2
}{\langle x^2(0)\rangle}~, \nonumber\\ V_p(t)=\frac{\langle
p^2(t)\rangle- {\langle p(t)\rangle}^2}{\langle p^2(0)\rangle}~.
\end{eqnarray}

\begin{figure}[htp]
 \scalebox{0.75}{\includegraphics{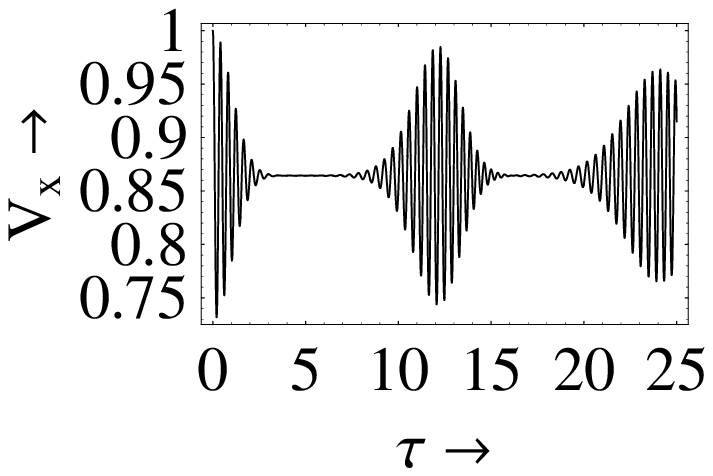}}% Here is how to import EPS art
 \caption{\label{Fig2(a)}Figure shows the variation of $V_x(t)$ as
a function of $\tau(\omega t=2 \pi \tau)$ with parameter
$\mu=0.1$. The average number $|\alpha|^2$ of excitation in the
QSM is 4.}
 \end{figure}

\begin{figure}[htp]
 \scalebox{0.75}{\includegraphics{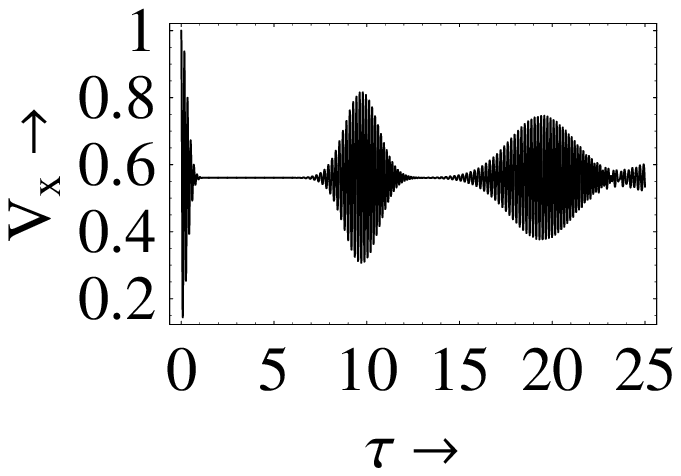}}% Here is how to import EPS art

\caption{\label{Fig2(a)}Figure shows the variation of $V_x(t)$ as
a function of $\tau(\omega t=2 \pi \tau)$ with parameter
$\mu=0.3$.  The average number $|\alpha|^2$ of excitation in the
QSM is 25.}
 \end{figure}

\noindent For the ground state of the oscillator

\begin{eqnarray}
\langle x(0)\rangle = \langle p(0)\rangle = 0,
\langle x p + p x \rangle = 0~, \nonumber\\
\langle x^2(0)\rangle =\frac{\hbar}{2 m \omega}, \langle
p^2(0)\rangle = \frac{m\hbar \omega}{2 }~.\label{eq25}
\end{eqnarray}

\noindent Using Eqs. (\ref{eq22}) and (\ref{eq25}) we get

\begin{eqnarray}
V_x(t) &= &
1-\sum_{n}\frac{|\alpha|^{2n}\exp[-|\alpha|^2]\sin^2[\omega t
\sqrt{1+\mu n}\hspace{0.8 mm}]}{n!}(\frac{\mu
n}{1+\mu n})~,\nonumber\\
\label{eq26}
\end{eqnarray}

\begin{eqnarray}
V_p(t) &= &
1+\sum_{n}\frac{|\alpha|^{2n}\exp[-|\alpha|^2]\sin^2[\omega t
\sqrt{1+\mu n}\hspace{0.8 mm}](\mu n)}{n!}~.\nonumber\\
\end{eqnarray}

\noindent Clearly $V_x(t)$ is always less than one and hence the
$x$ quadrature is squeezed. The corresponding result with a
classical source of modulation would be

\begin{eqnarray}
V_x(t) &= &1-\sin^2[\omega t\sqrt{1+\mu|\alpha|^2}\hspace{0.5
mm}](\frac{\mu |\alpha|^2}{1+\mu|\alpha|^2})~. \label{eq28}
\end{eqnarray}

\noindent The minimum value of $V_x$ in classical case is when
$\sin^2[\omega t\sqrt{1+\mu|\alpha|^2}]=1$ and hence

\begin{eqnarray}
 V_{x, min}=1-\big(\frac{\mu |\alpha|^2}{1+\mu|\alpha|^2}\big)~.
\end{eqnarray}

\noindent For the QSM the squeezing parameter (\ref{eq26})
exhibits typical collapse and revival of the classical periodical
oscillation [Figs. 3, 4]. This is again due to the discrete nature
of the quantum state of the source of modulation, i.e., control of
the spring constant by a quantum source. \vspace{1.5 mm}

\section{BACK REACTION OF THE OSCILLATOR ON QSM}
In this section we evaluate the back action of the oscillator on
the QSM. The Hamiltonian in  Eq. (\ref{eq2}) depends only on
$a^\dagger a$ and therefore only the off diagonal elements of the
density matrix $\rho_{QSM}$ are affected by the coupling of the
oscillator to QSM. The reduced state of the QSM is rather
involved. In order to appreciate the back action we examine the
mean displacement $\langle \hat{a} \rangle $ of the QSM which is
conditional on the measurement of oscillator in the state $
|\phi_{0} \rangle$. The mean value is given by

\begin{eqnarray}
d(t)&\equiv&\langle \hat{a}\rangle\nonumber\\
& = & \sum_{n,l} \langle l|a| n \rangle\langle n|\rho|l \rangle\nonumber\\
& = & \sum_{n}\sqrt{n}\rho_{n,n-1}~,\label{eq30}
\end{eqnarray}

\noindent where QSM's conditional density matrix is found to be

\begin{eqnarray}
\rho_{n,l}& = &\sum_{n,l}c_n^* c_l X_{n,l}~, \label{eq31}
\end{eqnarray}

\noindent with $X_{n,l}$ given by
\begin{eqnarray}
X_{n,l}& = &\langle \phi_{0}|\exp[i h(n)]\exp[-i
h(l)]|\phi_{0}\rangle~. \label{eq32}
\end{eqnarray}

\noindent In the above equation we have introduced the Hamiltonian
$h(n)$ defined by

\begin{eqnarray}
 h(n)& \equiv &
 {\frac{t}{\hbar}}(\frac{p^2}{2m}+\frac{m \omega^2(1+\mu
n)x^2}{2})~,
\end{eqnarray}

\noindent The ${X_{n,l}}$'s can be calculated by using operator
disentangling theorems for SU(1,1) group . We give details in the
Appendix(A). We find the following result[Eq. (\ref{eqA13})]

\begin{eqnarray}
X_{n,l} & = & \frac{(({\Gamma_{3n}})^*{\Gamma_{3
l}})^\frac{1}{4}}{\sqrt{1-(\Gamma_{+n})^* \Gamma_{+l}}}~.
\label{eq34}
\end{eqnarray}

\noindent The displacement of the QSM with $\Gamma$'s defined by
Eqs. (\ref{Gamma3}) and (\ref{Gammap}) can be obtained from Eqs.
(\ref{eq30}),(\ref{eq31}).

\begin{figure}[htp]
 \scalebox{0.85}{\includegraphics{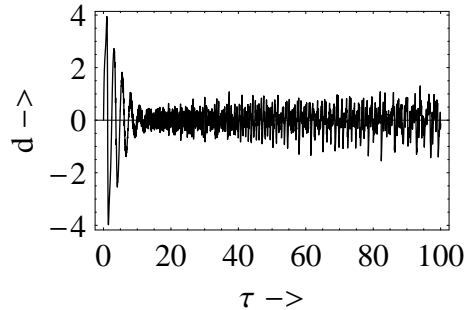}}% Here is how to import EPS art
 \caption{\label{Fig3}Figure shows the variation of
Imaginary$\langle \hat{a}\rangle$ as a function of $\tau(\omega
t=16 \pi \tau)$ with parameter $\mu=0.3$. The average number
$|\alpha|^2$ of excitation in the QSM is 25.}
 \end{figure}

In Fig. 5 we display the time evolution of the imaginary part of
$\langle \hat{a} \rangle $. Note that the imaginary part of
$\langle \hat{a} \rangle $ builds up starting from zero and
exhibits aperiodic behavior. The quantized nature of the optical
spring is well reflected in the back reaction \cite{naik}.
Evidently the back reaction occurs over a much larger time scale.

In conclusion we introduced the idea of a quantum optical spring ,
i.e., a spring whose spring constant is controlled by another
quantum system. The dynamics of the oscillator system exhibits the
phenomena of collapse and revivals including fractional revivals.
Further the spring gets entangled to the quantum system which
controls the spring constant.

\appendix
\section{CALCULATION OF THE MATRIX ELEMENT(\ref{eq32})}

In the appendix we give the details of our calculation for
$X_{nl}$. It is convenient to use $x$ and $p$ in terms of
anhilation and creation operators $\hat{b}$ and
${\hat{b}}^\dagger$

\begin{eqnarray}
&& x =
\sqrt{\frac{\hbar}{2m\omega}}({\hat{b}}^\dagger+\hat{b})~,\nonumber\\
&& p =  i \sqrt{\frac{\hbar
{m\omega}}{2}}({\hat{b}}^\dagger-\hat{b})~, \label{ihn}
\end{eqnarray}

\begin{eqnarray}
\exp[-i h(n)]|\phi_{0}\rangle = \exp[ -{\frac{i
t}{\hbar}}(\frac{p^2}{2m}+\frac{m \omega^2(1+\mu
n)x^2}{2})]|\phi_{0}\rangle~, \nonumber\\
\end{eqnarray}

\noindent Substituting the values for $x$ and  $p$ from Eq.
(\ref{ihn}) in the above equation and on simplification we get
\begin{eqnarray}
\exp[-i h(n)]|\phi_{0}\rangle = \exp[\gamma_{+
n}\hat{K_{+}}+\gamma_{-n}\hat{K_{-}}+\gamma_{3n}\hat{K_{3}}]|\phi_{0}\rangle~, \nonumber\\
\label{ihn1}
\end{eqnarray}

\noindent where

\begin{eqnarray}
\gamma_{\pm n} & \equiv & \frac{-i \mu\omega n t}{2}~,
\label{gaman}
\end{eqnarray}

\begin{eqnarray}
\gamma_{3n}&\equiv&-i \omega t(\mu n+2)~. \label{gaman1}
\end{eqnarray}

\begin{widetext}

\noindent In (\ref{ihn1}) $\hat{K_\pm}$, $\hat{K_3}$ are the
generators of SU(1,1) algebra given by

\begin{eqnarray}
\hat{K}_{3} &\equiv& \frac{1}{4}({\hat{b}}^\dagger\hat{b}+\hat{b}
{\hat{b}}^\dagger)~,
\end{eqnarray}

\begin{eqnarray}
\hat{K}_{+}& =
&(\hat{K}_-)^\dag=\frac{1}{2}({\hat{b}}^\dagger)^2~.
\end{eqnarray}

\noindent The Eq. (\ref{ihn1}) can be simplified by using the
disentangling theorem \cite{barnett} for the SU(1,1) group:

\begin{eqnarray}
\exp(\gamma_{+ n}\hat{K}_{+}+\gamma_{- n}\hat{K_{-}}+\gamma_{3
n}\hat{K_{3}})= \exp(\Gamma_{+ n}\hat{K_{+}})\exp[(\ln \Gamma_{3
n})\hat{K_{3}}] \exp(\Gamma_{-
n}\hat{K_{-}})~,\label{disentangling}
\end{eqnarray}
\noindent where

\begin{eqnarray}
\Gamma_{3n}&\equiv& \frac{1}{(\cosh \beta_{n}-\frac{\gamma_{3
n}}{2\beta_{n}}\sinh\beta_{n})^{2}}~, \label{Gamma3}
\end{eqnarray}

\begin{eqnarray}
\Gamma_{\pm n}&\equiv& \frac{2\gamma_{\pm
n}\sinh\beta_{n}}{(2\beta_{n} \cosh \beta_{n}-\gamma_{3
n}\sinh\beta_{n})}~,\label{Gammap}
\end{eqnarray}

\begin{eqnarray}
{\beta_{n}}^2&\equiv&
\frac{1}{4}{\gamma_{3n}}^{2}-\gamma_{+n}\gamma_{-n}~.
\end{eqnarray}

{\noindent}We use the disentangling theorem in the form
(\ref{disentangling}) as then $\hat{K_{-}}$ acting on
$|\phi_{0}\rangle$ yields zero and $\hat{K_{3}}$ terms can be
simplified as $|\phi_{0}\rangle$ is an eigenstate of
$\hat{K_{3}}$. Thus we reduce Eq. (\ref{ihn1}) to
\begin{eqnarray}
\exp(\gamma_{+n}\hat{K_{+}}+\gamma_{-n}\hat{K_{-}}+\gamma_{3n}\hat{K_{3}})|\phi_{0}\rangle
&\equiv& \exp(\Gamma_{+n} \hat{K_{+}})
\exp[(\ln\Gamma_{3n})\hat{K_{3}}]
\exp(\Gamma_{-n}\hat{K_{-}})|\phi_{0}\rangle\nonumber\\
&=&\exp[\frac{(\ln\Gamma_{3n})}{4}]
\exp[\Gamma_{+n} \hat{K_{+}}]|\phi_{0}\rangle \nonumber\\
&=&{(\Gamma_{3n})}^\frac{1}{4} \exp(\Gamma_{+n}
\hat{K_{+}})|\phi_{0}\rangle
\end{eqnarray}

Hence $X_{n,l}$ as defined by (\ref{eq32}) can be calculated as
follows

\begin{eqnarray}
X_{n,l}& = &\langle\phi_{0}|\exp(i h(n))\exp(- i h(l))|\phi_{0}\rangle \nonumber\\
& =
&(({\Gamma_{3n}})^*{\Gamma_{3l}})^\frac{1}{4}\langle\phi_{0}|\exp((\Gamma_{+n})^*
\hat{K_{-}})\exp(\Gamma_{+l}
\hat{K_{+}})|\phi_{0}\rangle \nonumber\\
& =
& (({\Gamma_{3n}})^*{\Gamma_{3 l}})^\frac{1}{4}\sum_{p}\frac{{((\Gamma_{+n})^* \Gamma_{+l})}^p ({2p})!}{({p}!)^2 2^{2p}}\nonumber\\
& = &
\frac{(({\Gamma_{3n}})^*{\Gamma_{3l}})^\frac{1}{4}}{\sqrt{1-(\Gamma_{+n})^*
\Gamma_{+l}}}~.\label{eqA13}
\end{eqnarray}
This is the result we use in Section V to calculate the
back-action.

\vspace{1.5 mm}

\end{widetext}


\begin{thebibliography}{999}
\bibitem{footnote1}We do not
include damping effects in the analysis which are known to be
detrimental for the observation of such revival effects.
\end{thebibliography}
\end{document}